\documentclass[aps,prc,groupedaddress,showpacs,manuscript]{revtex4-1}
\usepackage{graphicx}% Include figure files

\usepackage{colordvi}

\begin{document}
%\begin{CJK*} {UTF8}{} %{GB} {gbsn}
%\preprint{APS/123-QED}

\title{Helium-3 production from Pb+Pb collisions at SPS energies with the UrQMD model and the traditional coalescence afterburner}

\author {Qingfeng Li$\, ^{1}$\footnote{E-mail address: liqf@hutc.zj.cn},
Yongjia Wang$\, ^{1}$,
%\footnote{E-mail address: lzuwyj@hotmail.com},
Xiaobao Wang$\, ^{1}$,
and Caiwan Shen$\, ^{1}$,
}

\affiliation{
1) School of Science, Huzhou University, Huzhou 313000, P.R. China \\
 }
\date{\today}

\begin{abstract}

A potential version of the UrQMD (UrQMD/M) transport model and a traditional coalescence model are combined to calculate the production of $^3$He fragments in central Pb+Pb collisions at SPS energies 20-80 GeV/nucleon.  It is found that the Lorentz transformation in the afterburner influences visibly the $^3$He yield and should be considered in calculations. The rapidity distribution of $^3$He multiplicities (including the concave shape) can be described well with UrQMD/M when it stops during t$_{\rm cut}$=100$\pm$25 fm$/c$ and the coalescence afterburner with one parameter set of ($R_0$,$P_0$)=(3.8 fm, 0.3 GeV$/$c) is taken into use afterwards.

\end{abstract}



\maketitle
%\end{CJK*}

\section{Motivation and model settings}
The production and decay properties of nuclei are fundamental many-body problems in the evolution of the universe, which can be studied in artificial laboratories with heavy ion collisions (HICs). Besides the HICs at low and intermediate energies for the synthesis of (super-)heavy nuclei or the multi-fragmentation through a possible liquid gas phase transition, the production mechanism of nuclei in ultra-relativistic HICs deserves more investigation since it may give important message on the quantum chromodynamics (QCD) phase transition from quark-gluon plasma (QGP) to hadron gas (HG) \cite{Arsenescu:2003eg}. In addition, if nuclei are formed through a coalescence process one can use their yield ratios to measure the volume of the particle source from which they emerge. This offers important information on the space-time evolution of the reaction, and implies a close relation of space-time structure between the coalescence and the the so-called ``femtoscopy'' or ``HBT'' (in reference to Hanbury-Brown and Twiss¡¯s original work with photons) correlation \cite{Anticic:2004yj}.

In past two decades, profited from some experimental measurements e.g., \cite{Johnson:1997gh,Anticic:2004yj,Chekanov:2007mv,Anticic:2011ny,Newman:2013ada}, the light fragment production mechanism has being investigated more deeply but mainly with a coalescence afterburner in which a Wigner-function method is in use (called Wigner-coalescence) \cite{Mattiello:1995xg,Nagle:1996vp,Monreal:1999mv}. However, it has the substantial disadvantage of not conserving baryon number in the projection. Therefore, we employ the traditional phase-space coalescence approach frequently used for HICs at low and intermediate energies \cite{Kruse:1985pg,Li:2005kqa,Wang:2013wca} and in this paper as well. In Ref.~\cite{Li:2015aa}, the method has been used for describing rapidity distributions of both the E895 proton data at AGS energies and the NA49 net proton data at SPS energies with the help of the Ultra-relativistic Quantum Molecular Dynamics (UrQMD) model supplemented by potentials for both pre-formed hadrons and confined baryons (called UrQMD/M) \cite{Li:2007yd,Li:2008ge}, taking the similar density-dependent (Skyrme-like) terms. It is found that, using only one parameter set of ($R_0$, $P_0$)=(3.8 fm, 0.3 GeV$/$c) in the afterburener (where $R_0$ and $P_0$ are parameters of relative distance and relative momentum between two particles for constructing clusters), both sets of experimental data can be described reasonably well. This success encourages us to examine further the production yields of light fragments from HICs at high energies. It is noticed that some experimental data related to the production of light clusters such as deuterons, tritons, and $^3$He have been available \cite{Anticic:2004yj,Blume:2007kw,Kolesnikov:2007ps}.  In this paper, the $^3$He production from central Pb+Pb reactions at SPS energies 20-80 GeV/nucleon is taken as an example.

The UrQMD microscopic transport model was originally developed to study particle production at high energies such as AGS, SPS, and RHIC energies \cite{Bass98,Bleicher99,urqmdhomepage}. Recently, it has been updated for simulating HICs at both lower, such as SIS energies \cite{Li:2011zzp,Guo:2012aa,Wang:2012sy,Wang:2013wca,Wang:2014aba} and higher, such as LHC energies \cite{Li:2012ta,Graef:2012za,Graef:2012sh}. It is interesting to see that the potentials always play an important role on the particle emission from HICs at whatever low or high energies. Especially, with the consideration of mean-field potentials for pre-formed hadrons, some quantities such as the HBT of two particles
(especially the time-related HBT-puzzle), the elliptic flow (in the cascade mode calculations, it is known as a flow-puzzle), and the yields of strange baryons or anti-baryons (a puzzle related to the strangeness enhancement) can be better described or explained \cite{Li:2007yd,Li:2008ge,Li:2010ie}. Although a thorough explanation of all existing ¡°puzzles¡± is still awaiting since a complete description of the multi-particle collision dynamics crossing a possible phase transition and/or a consistency with the first-principle lattice QCD calculations has not arrived yet, the current version of UrQMD (UrQMD/M) is nice for the investigation of the light fragment production mechanism if a suitable afterburner is linked when the UrQMD stops at a certain time t$_{\rm cut}$.

\begin{figure}[htbp]
\centering
\includegraphics[angle=0,width=0.8\textwidth]{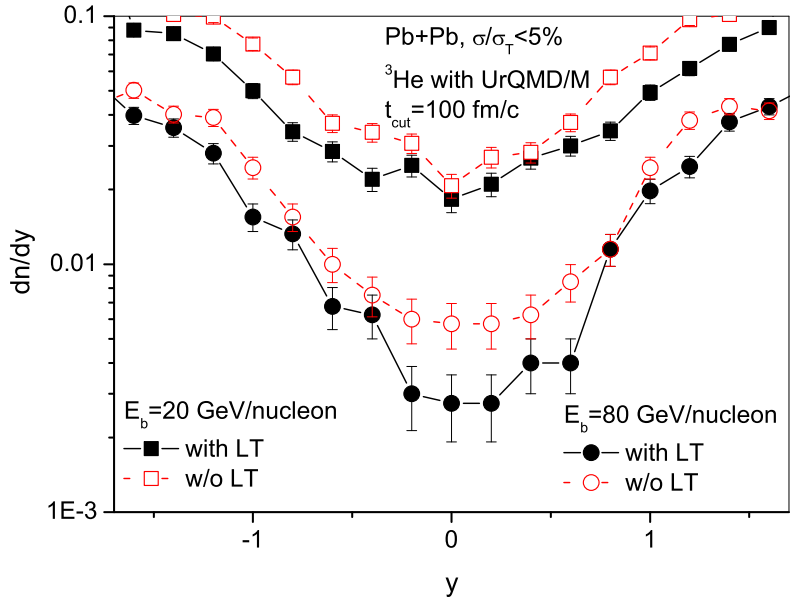}
\caption{\label{fig1} (Color online) Rapidity distribution of $^3$He from central Pb+Pb collisions with beam energies $E_b=20$ (lines with squares) and $80$ GeV$/$nucleon (lines with circles). t$_{\rm cut}$=100 fm$/c$ is chosen. For each beam energy, results with (solid symbols) and without (open symbols) LT are shown.
}
\end{figure}

In the afterburner, as stated in Ref.~\cite{Li:2015aa} for protons, the relativistic effect ought to be examined when calculating relative distance $\delta r$ and relative momentum $\delta p$ between two baryons. It was found that, due to the large cancellation between the coordinate-spatial expansion and the momentum-spatial shrinkage by the Lorentz transformation (LT) in the afterburner, the proton yield with LT is close to that without LT, although some minor difference between them is still observed. It is interesting to see how the minor difference in proton yield influences the yield of light fragment such as $^3$He. Fig.~\ref{fig1} depicts the rapidity $y$ ($=\frac{1}{2}\mathrm{log}(\frac{E_{\mathrm{cm}}+p_{//}}{E_{\mathrm{cm}}-p_{//}})$, where $E_{\mathrm{cm}}$ and $p_{//}$ are the energy and longitudinal momentum of the observed particle in the center-of-mass system, respectively) distribution of $^3$He from central ($<5\%$ of the total cross section $\sigma_T$) Pb+Pb collisions with beam energies $E_b=20$ and $80$ GeV$/$nucleon. t$_{\rm cut}$=100 fm$/c$ is chosen. For each beam energy, results with and without consideration of LT are shown for comparison. It is clear that the influence of LT is larger for light fragments than for free nucleons, which is mainly due to their much smaller yields. In addition, the consideration of LT in the afterburner drives further down the yield of $^3$He, which is attributed to the larger relativistic effect for the momentum difference $\delta p$ than for the distance difference $\delta r$, as discussed in Ref.~\cite{Li:2015aa}. Hence, in the following discussions, the LT effect is always taken into account in the coalescence model.

\section{Time dependence of $^3$He multiplicities}

\begin{figure}[htbp]
\centering
\includegraphics[angle=0,width=0.7\textwidth]{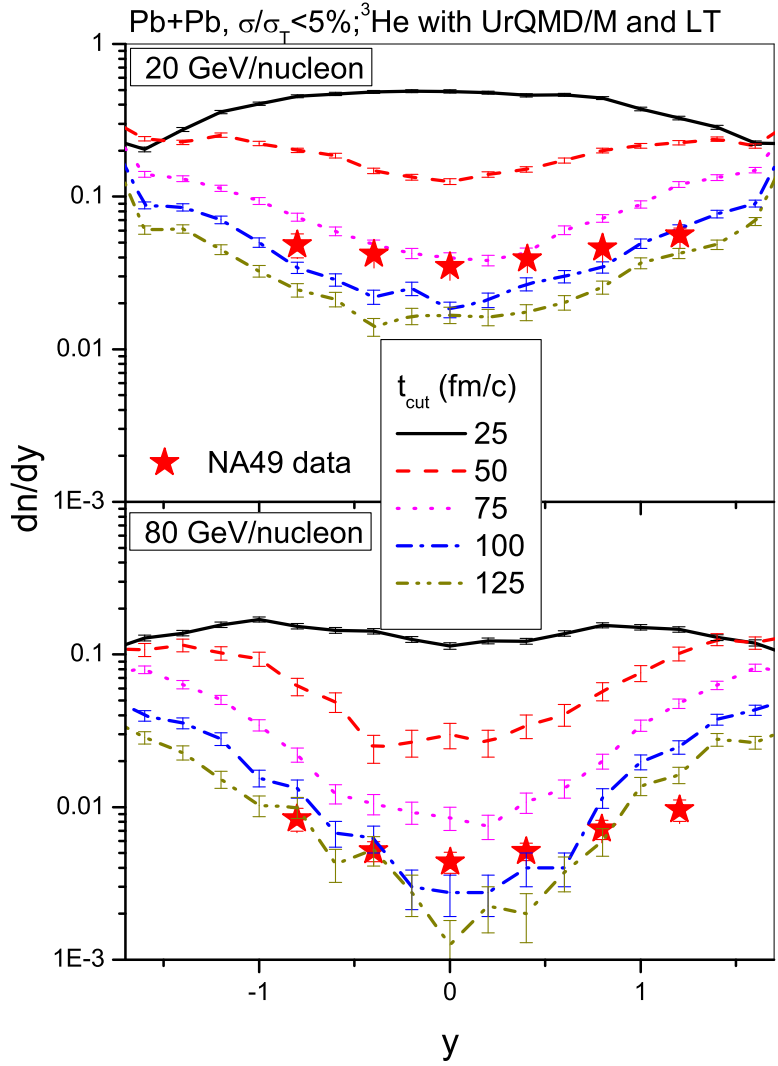}
\caption{\label{fig2} (Color online) Rapidity distribution of $^3$He multiplicities at two beam energy points: $20$ (top plot) and $80$ GeV$/$nucleon (bottom plot). In each plot, results at t$_{\rm cut}$=25 (solid line), 50 (dashed line), 75 (dotted line), 100 (dash-dotted line), and 125 fm$/c$ (dash-dot-dotted line) are compared to the NA49 experimental data taken from Ref.~\cite{Blume:2007kw}.}
\end{figure}

For a systematic survey, the same parameter set ($R_0$, $P_0$)=(3.8 fm, 0.3 GeV$/$c) in the phase-space coalescence model used for protons is still adopted for the current calculations. But, the t$_{\rm cut}$ dependence of the $^3$He multiplicity should be addressed since it might partly produced later than protons due to the sequential decay of highly-excited heavier fragments. In Fig.~\ref{fig2} we demonstrate the rapidity distribution of $^3$He multiplicities at two beam energy points $20$ (top plot) and $80$ GeV$/$nucleon (bottom plot) and at several stopping times 25, 50, 75, 100, and 125 fm$/c$ (different lines), respectively. Correspondingly, the NA49 experimental data are taken from Ref.~\cite{Blume:2007kw} (scattered stars). At t$_{\rm cut}$=25 fm$/c$, which is known that the high-density compression phase has disappeared for a long time \cite{Li:2010ie}, a large amount of nucleons and light fragments are frozen out due to the following low-density environment, which is seen with the solid lines at both beam energies. However, it is clear that at this time the multiplicities of $^3$He are at least one order of magnitude larger than experimental data. In addition, the concave shape shown in data can not be described by calculations. It implies that most of the $^3$He ``constructed'' at the early times will decay to lighter clusters such as deuterons and nucleons. With the time increasing from 25 to 75 fm$/c$, it is seen clearly that the $^3$He multiplicities are reduced quickly and approach to the data. Meanwhile, the concave shape emerges. Due to larger stopping and more energy deposition, more excited $^3$He fragments at mid-rapidity decay than those at projectile-target rapidities. As the time increases further, the $^3$He multiplicities is seen to reduce continuously but with much lower speed. Due to the lack of a statistic treatment for the very late stage, the NA49 data are described well enough within the stopping time t$_{\rm cut}$=100$\pm$25 fm$/c$ for UrQMD/M together with the phase-space coalescence model using the parameter set ($R_0$, $P_0$)=(3.8 fm, 0.3 GeV$/$c). One also finds that it takes longer time for $^3$He fragments lying away from mid-rapidity to meet the data which is due to the fact that much more heavier fragments exist in these areas and decay to lighter ones such as $^3$He.

\begin{figure}[htbp]
\centering
\includegraphics[angle=0,width=0.8\textwidth]{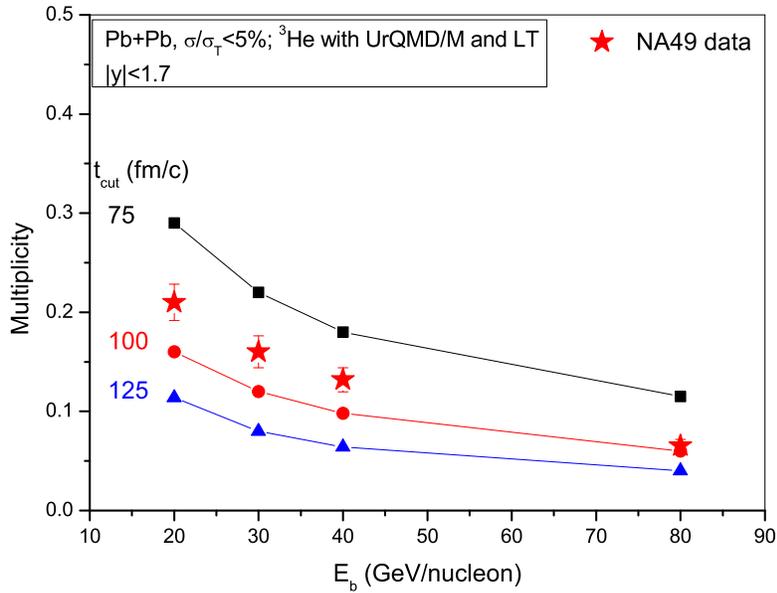}
\caption{\label{fig3} (Color online) Rapidity-integrated ($|y|<1.7$) multiplicities of $^3$He fragments as a function of beam energy, calculated with stopping times 75(line with squares), 100 (line with circles), and 125 fm$/c$ (line with up-triangles) for UrQMD/M. The NA49 data are taken from Ref.~\cite{Blume:2007kw}.
}
\end{figure}

This time dependence can also been seen from the rapidity-integrated ($|y|<1.7$) multiplicity of $^3$He fragments which is shown in Fig.~\ref{fig3}, as a function of SPS beam energies. Calculations with t$_{\rm cut}$=75, 100, and 125 fm$/c$ (lines with different symbols) are shown for comparison to the NA49 data \cite{Blume:2007kw} (scattered stars). With the increase of beam energy, the calculated $^3$He yield decreases and follows data reasonable well. Further, it is interesting to see that it requires a longer time to reach the experimental data of light fragments at higher beam energies. At first glance, it is hard to understand since higher excitation should lead to earlier emission. It is true if we take a look at the decrease of absolute values at one fixed time with beam energies increasing from 20 to 80 GeV$/$nucleon, as seen in Fig.~\ref{fig2}. However, meanwhile, the higher energy of heavier fragments leads to the more sequential decay to lighter fragments which will certainly take a longer time. Therefore, the competition between production and decay of excited $^3$He fragments at different rapidities determines the quantity and time scale of its final production, during which a proper dynamic treatment is obviously important since it determines the phase-space evolution and the final stability of a fragment. Meanwhile, the production of clusters other than $^3$He should be also investigated in a systematic manner so that a more complete prospect could be established, which is in progress.

\section{Summary}
In summary, with a potential version of the UrQMD (UrQMD/M) transport model, and a traditional coalescence model in which one parameter set of ($R_0$,$P_0$)=(3.8 fm, 0.3 GeV$/$c) is used, both the rapidity distribution and the rapidity-integrated multiplicity of $^3$He fragments in central Pb+Pb collisions at SPS energies 20-80 GeV/nucleon are calculated at several stopping times of the transport program, and with or without considering the Lorentz transformation in the afterburner. It is found that the Lorentz transformation influences visibly the $^3$He yield and should be considered in the analysis. The rapidity distribution of $^3$He multiplicities (including the concave shape) can be described well with UrQMD/M when it stops during t$_{\rm cut}$=100$\pm$25 fm$/c$ and the coalescence afterburner is linked together. The finding that the $^3$He cluster can only escape at the later freeze-out stage within a dilute environment supports the production mechanism of light fragment via coalescence as well. The universal competition between sequential production and decay of light fragments asks us to consider more carefully both the stiffness of the dynamic evolution and the statistical sequential decay.

\begin{acknowledgements}
We thank Prof. M. Bleicher for valuable suggestions and acknowledge support by the computing server C3S2 in Huzhou
University. The work is supported in part by the National
Natural Science Foundation of China (Nos. 11375062, 11275068, 11505056, and 11505057), the project sponsored by SRF for ROCS, SEM, the Education Bureau of Zhejiang Province (Y201533176) and the Doctoral Scientific Research Foundation (No. 11447109).
\end{acknowledgements}


\begin{thebibliography}{0}

%\cite{Arsenescu:2003eg}
\bibitem{Arsenescu:2003eg}
  R.~Arsenescu {\it et al.} [NA52 Collaboration],
  %``An Investigation of the anti-nuclei and nuclei production mechanism in Pb + Pb collisions at 158-A-GeV,''
  New J.\ Phys.\  {\bf 5}, 150 (2003).
  %%CITATION = NJOPF,5,150;%%
  %4 citations counted in INSPIRE as of 31 Aug 2015

%\cite{Anticic:2004yj}
\bibitem{Anticic:2004yj}
  T.~Anticic {\it et al.} [NA49 Collaboration],
  %``Energy and centrality dependence of deuteron and proton production in Pb + Pb collisions at relativistic energies,''
  Phys.\ Rev.\ C {\bf 69}, 024902 (2004).
  %%CITATION = PHRVA,C69,024902;%%
  %96 citations counted in INSPIRE as of 31 Aug 2015

%\cite{Johnson:1997gh}
\bibitem{Johnson:1997gh}
  S.~C.~Johnson,
  %``Light fragment formation in 11.5-GeV/c per nucleon Au + Au collisions,''
  UMI-98-22959.
  %%CITATION = UMI-98-22959;%%

%\cite{Chekanov:2007mv}
\bibitem{Chekanov:2007mv}
  S.~Chekanov {\it et al.} [ZEUS Collaboration],
  %``Measurement of (anti)deuteron and (anti)proton production in DIS at HERA,''
  Nucl.\ Phys.\ B {\bf 786}, 181 (2007)
  [arXiv:0705.3770 [hep-ex]].
  %%CITATION = ARXIV:0705.3770;%%
  %15 citations counted in INSPIRE as of 31 Aug 2015


%\cite{Anticic:2011ny}
\bibitem{Anticic:2011ny}
  T.~Anticic {\it et al.} [NA49 Collaboration],
  %``Antideuteron and deuteron production in mid-central Pb+Pb collisions at 158$A$ GeV,''
  Phys.\ Rev.\ C {\bf 85}, 044913 (2012)
  [arXiv:1111.2588 [nucl-ex]].
  %%CITATION = ARXIV:1111.2588;%%
  %5 citations counted in INSPIRE as of 31 Aug 2015

%\cite{Newman:2013ada}
\bibitem{Newman:2013ada}
  P.~Newman and M.~Wing,
  %``The Hadronic Final State at HERA,''
  Rev.\ Mod.\ Phys.\  {\bf 86}, no. 3, 1037 (2014)
  [arXiv:1308.3368 [hep-ex]].
  %%CITATION = ARXIV:1308.3368;%%
  %6 citations counted in INSPIRE as of 31 Aug 2015


%\cite{Mattiello:1995xg}
\bibitem{Mattiello:1995xg}
  R.~Mattiello, A.~Jahns, H.~Sorge, H.~Stoecker and W.~Greiner,
  %``Deuteron flow in ultrarelativistic heavy ion reactions,''
  Phys.\ Rev.\ Lett.\  {\bf 74}, 2180 (1995).
  %%CITATION = PRLTA,74,2180;%%
  %64 citations counted in INSPIRE as of 09 juil. 2015

%\cite{Nagle:1996vp}
\bibitem{Nagle:1996vp}
  J.~L.~Nagle, B.~S.~Kumar, D.~Kusnezov, H.~Sorge and R.~Mattiello,
  %``Coalescence of deuterons in relativistic heavy ion collisions,''
  Phys.\ Rev.\ C {\bf 53}, 367 (1996).
  %%CITATION = PHRVA,C53,367;%%
  %49 citations counted in INSPIRE as of 09 juil. 2015

%\cite{Monreal:1999mv}
\bibitem{Monreal:1999mv}
  B.~Monreal, S.~A.~Bass, M.~Bleicher, S.~Esumi, W.~Greiner, Q.~Li, H.~Liu and W.~J.~Llope {\it et al.},
  %``Deuterons and space momentum correlations in high-energy nuclear collisions,''
  Phys.\ Rev.\ C {\bf 60}, 031901 (1999)
  [nucl-th/9904080].
  %%CITATION = NUCL-TH/9904080;%%
  %10 citations counted in INSPIRE as of 09 juil. 2015


%\cite{Kruse:1985pg}
\bibitem{Kruse:1985pg}
  H.~Kruse, B.~V.~Jacak, J.~J.~Molitoris, G.~D.~Westfall and H.~Stoecker,
  %``Vlasov-Uehling-Uhlenbeck Theory Of Medium-Energy Heavy Ion Reactions: Role
  %Of Mean Field Dynamics And Two-Body Collisions,''
  Phys.\ Rev.\  C {\bf 31}, 1770 (1985).
  %%CITATION = PHRVA,C31,1770;%%

%\cite{Li:2005kqa}
\bibitem{Li:2005kqa}
  Q.~Li, Z.~Li, S.~Soff, M.~Bleicher and H.~Stoecker,
  %``Probing the density dependence of the symmetry potential at low and high densities,''
  Phys.\ Rev.\ C {\bf 72}, 034613 (2005)
  [nucl-th/0506030].
  %%CITATION = NUCL-TH/0506030;%%
  %48 citations counted in INSPIRE as of 28 May 2015

%\cite{Wang:2013wca}
\bibitem{Wang:2013wca}
  Y.~Wang, C.~Guo, Q.~Li, H.~Zhang, Z.~Li and W.~Trautmann,
  %``Collective flows of light particles in the Au+Au collisions at intermediate energies,''
  Phys.\ Rev.\ C {\bf 89}, 034606 (2014)
  [arXiv:1305.4730 [nucl-th]].
  %%CITATION = ARXIV:1305.4730;%%
  %7 citations counted in INSPIRE as of 28 May 2015

\bibitem{Li:2015aa}
  Q.~Li, Y.~Wang, X.~Wang, and C.~Shen,
  Sci.\ China Phys.\ Mech.\ Astron.\  {\bf ??}, ?? (??).


%\cite{Li:2007yd}
\bibitem{Li:2007yd}
  Q.~Li, M.~Bleicher and H.~Stocker,
  %``The Effect of pre-formed hadron potentials on the dynamics of heavy ion collisions and the HBT puzzle,''
  Phys.\ Lett.\ B {\bf 659}, 525 (2008)
  [arXiv:0709.1409 [nucl-th]].
  %%CITATION = ARXIV:0709.1409;%%
  %34 citations counted in INSPIRE as of 26 May 2015

%\cite{Li:2008ge}
\bibitem{Li:2008ge}
  Q.~Li, M.~Bleicher and H.~Stocker,
  %``Transport model study of the m(T)-scaling for Lambda, K, and pi HBT-correlations,''
  Phys.\ Lett.\ B {\bf 663}, 395 (2008)
  [arXiv:0802.3618 [nucl-th]].
  %%CITATION = ARXIV:0802.3618;%%
  %13 citations counted in INSPIRE as of 31 Aug 2015

%\cite{Blume:2007kw}
\bibitem{Blume:2007kw}
  C.~Blume [Na49 Collaboration],
  %``Centrality and energy dependence of proton, light fragment and hyperon production,''
  J.\ Phys.\ G {\bf 34}, S951 (2007)
  [nucl-ex/0701042].
  %%CITATION = NUCL-EX/0701042;%%
  %21 citations counted in INSPIRE as of 31 Aug 2015
  
%\cite{Kolesnikov:2007ps}
\bibitem{Kolesnikov:2007ps}
  V.~I.~Kolesnikov [NA49 Collaboration],
  %``Anti-nuclei and nuclei production in Pb+Pb collisions at CERN SPS energies,''
  J.\ Phys.\ Conf.\ Ser.\  {\bf 110}, 032010 (2008)
  [arXiv:0710.5118 [nucl-ex]].
  %%CITATION = ARXIV:0710.5118;%%
  %6 citations counted in INSPIRE as of 15 Oct 2015
  

\bibitem{Bass98}S. A. Bass {\it et al.}, [UrQMD-Collaboration], Prog. Part. Nucl. Phys. {\bf 41}, 255 (1998).

\bibitem{Bleicher99}M. Bleicher {\it et al.}, [UrQMD-Collaboration], J. Phys. G: Nucl.
Part. Phys. {\bf 25}, 1859 (1999).

\bibitem{urqmdhomepage} see UrQMD homepage, www.urqmd.org.

%\cite{Li:2011zzp}
\bibitem{Li:2011zzp}
  Q.~Li, C.~Shen, C.~Guo, Y.~Wang, Z.~Li, J.~Lukasik and W.~Trautmann,
  %``Nonequilibrium dynamics in heavy-ion collisions at low energies available at the GSI Schwerionen Synchrotron,''
  Phys.\ Rev.\ C {\bf 83}, 044617 (2011).
  %%CITATION = PHRVA,C83,044617;%%
  %17 citations counted in INSPIRE as of 31 Aug 2015

%\cite{Guo:2012aa}
\bibitem{Guo:2012aa}
 C.~Guo,  Y.~Wang, Q.~Li, W.~Trautmann, L.~Liu and L.~Wu,
  Sci.\ China Phys.\ Mech.\ Astron.\  {\bf 55}, 252 (2012).

%\cite{Wang:2012sy}
\bibitem{Wang:2012sy}
  Y.~Wang, C.~Guo, Q.~Li and H.~Zhang,
  %``The effect of symmetry potential on the balance energy of light particles emitted from mass symmetric heavy-ion collisions with isotopes, isobars and isotones,''
  Sci.\ China Phys.\ Mech.\ Astron.\  {\bf 55}, 2407 (2012).
  %%CITATION = 00765,55,2407;%%
  %4 citations counted in INSPIRE as of 31 Aug 2015


%\cite{Wang:2014aba}
\bibitem{Wang:2014aba}
  Y.~Wang, C.~Guo, Q.~Li and H.~Zhang,
  %``$^{3}$H/$^{3}$He ratio as a probe of the nuclear symmetry energy at sub-saturation densities,''
  Eur.\ Phys.\ J.\ A {\bf 51}, 37 (2015)
  [arXiv:1407.7625 [nucl-th]].
  %%CITATION = ARXIV:1407.7625;%%

%\cite{Li:2012ta}
\bibitem{Li:2012ta}
  Q.~Li, G.~Graf and M.~Bleicher,
  %``UrQMD calculations of two-pion HBT correlations in central Pb-Pb collisions at $\sqrt{s_{NN}}=2.76$ TeV,''
  Phys.\ Rev.\ C {\bf 85}, 034908 (2012)
  [arXiv:1203.4104 [nucl-th]].
  %%CITATION = ARXIV:1203.4104;%%
  %13 citations counted in INSPIRE as of 31 Aug 2015

%\cite{Graef:2012za}
\bibitem{Graef:2012za}
  G.~Graef, Q.~Li and M.~Bleicher,
  %``Formation time dependence of femtoscopic $\pi \pi$ correlations in p+p collisions at $\sqrt{s_{NN}}$=7 TeV,''
  J.\ Phys.\ G {\bf 39}, 065101 (2012)
  [arXiv:1203.4421 [nucl-th]].
  %%CITATION = ARXIV:1203.4421;%%
  %3 citations counted in INSPIRE as of 31 Aug 2015

%\cite{Graef:2012sh}
\bibitem{Graef:2012sh}
  G.~Graef, M.~Bleicher and Q.~Li,
  %``Examination of scaling of Hanbury-Brown--Twiss radii with charged particle multiplicity,''
  Phys.\ Rev.\ C {\bf 85}, 044901 (2012)
  [arXiv:1203.4071 [nucl-th]].
  %%CITATION = ARXIV:1203.4071;%%
  %8 citations counted in INSPIRE as of 31 Aug 2015



%\cite{Li:2010ie}
\bibitem{Li:2010ie}
  Q.~Li and Z.~Li,
  %``Production and rescattering of strange baryons at SPS energies in a transport model with hadron potentials,''
  Mod.\ Phys.\ Lett.\ A {\bf 27}, 1250004 (2012)
  [arXiv:1010.2570 [nucl-th]].
  %%CITATION = ARXIV:1010.2570;%%

\end{thebibliography}
\end{document}